\documentclass[useAMS,usenatbib]{mnras}
\usepackage{epsfig}
\usepackage{amssymb}
\usepackage{upgreek}

\newcommand\ion[2]{#1\,{\scshape{#2}}}

\title[Dust in broad absorption line quasars] {Attenuation from the optical to the extreme ultraviolet by dust associated with broad absorption line quasars: the driving force for outflows}
\author[C. M. Gaskell et al.]{C. Martin Gaskell$^{1}$\thanks{E-mail:
mgaskell@ucsc.edu}, Jake J. M. Gill$^{1}$, Japneet Singh$^{1,2}$ \\
\\$^1$Department of Astronomy and Astrophysics, University of California, Santa Cruz, CA 95064
\\$^2$Department of Computer Science, University of Illinois Urbana-Champaign, Urbana, IL 61801}

\begin{document}

\date{Received 2023 November 7; in original form 2016 November 11}

\pagerange{\pageref{firstpage}--\pageref{lastpage}} \pubyear{2020}

\maketitle

\label{firstpage}

\begin{abstract}
We derive a mean attenuation curve out to the rest-frame extreme ultraviolet (EUV) for ‘BAL dust’ – the dust causing the additional extinction of active galactic nuclei (AGNs) with broad absorption lines (BALQSOs). In contrast to the normal, relatively-flat, mean AGN attenuation curve, BAL dust is well fit by a steeply-rising, SMC-like curve.  We confirm the shape of the theoretical Weingartner \& Draine SMC curve out to 700 Angstroms but the drop in attenuation at still shorter wavelengths is less than predicted. { The similar SMC-like attenuation curve for low-ionization BALQSOs (LoBALs) does not support the idea that they are an early phase in the life of an AGN when it is breaking out of a cocoon of star-forming dust.} Although the attenuation is only $E(B - V) \sim 0.03 - 0.05$ in the optical, it rises to one magnitude in the EUV, which is an optimum value for radiative acceleration of dusty gas. Because the spectral energy distribution of AGNs peaks in the EUV, the force on the dust dominates the acceleration of BAL gas. { Although the shape of the attenuation curve for LoBALs is similar to the shape for HiBALs, the LoBALs on average show negative attenuation in the optical.} This is naturally explained if there is more light scattered into our line of sight in LoBALs compared with non-BALQSOs. We suggest that this and partial covering are causes when attenuation curves appear to be steeper in the UV than an SMC curve.

\end{abstract}


\section{Introduction}

Broad absorption lines (BALs) have been known in active galactic nuclei (AGNs) for a long time \citep{Lynds67}, but the cause of the very high velocity outflows in broad absorption line AGNs (BALQSOs) remains poorly understood.  It was initially widely assumed that there was no dust associated with the BAL gas.  However, \citet{Weymann+91} noticed that while BALQSOs showing high-ionization BALs (so called HiBALs) appeared to have { generally} similar { emission-line properties and continuum shapes} as non-BALQSOs, the subset of BALQSOs showing low-ionization BALs (LoBALs) appeared to have significantly redder UV continua.  They also { suspected} that the UV continua of HiBALs might be somewhat redder than the continua of non-BALQSOs. \citet{Sprayberry+Foltz92} showed that the differences in the spectral energy distributions of LoBALs and HiBALs could be explained by greater extinction in the LoBALs so long as the reddening curve lacked the 2175 \AA\ absorption feature. At that time the only known reddening curve lacking 2175 \AA~absorption was the extinction curve for the Small Magellanic Cloudy (SMC) and this
gave an excess reddening for LoBALs of $E(B-V) \sim 0.1$. \citet{Yamoto+Vansevicius99} showed that not only LoBALs but also HiBALs were reddened compared with non-BALQSOs.  Considering only radio-loud AGNs and assuming an SMC curve, \citet{Brotherton+01} found mean reddenings of $E(B-V) \sim 0.04$ and $\sim 0.10$ for HiBALs and LoBALs respectively relative to non-BALQSOs. \citet{Reichard+03}, again assuming an SMC reddening curve, similarly found mean reddenings of $E(B-V) \sim 0.02$ and $\sim 0.08$ respectively for optically-selected HiBALs and LoBALS from the SDSS.  The somewhat lower mean excess reddenings for the SDSS BALQSOs are expected since, as \citet{Reichard+03} note, the sample of \citet{Brotherton+01} was selected at longer wavelengths and is thus more robust against losing reddened AGNs from the sample.

Following \citet{Sprayberry+Foltz92}, studies of the excess extinction associated with BALQSOs have assumed that the extinction curve of the dust is an SMC-like curve because of the lack of $\lambda$2175 absorption.  However, subsequent work has shown that there are {\em other} extinction curves lacking $\lambda$2175 absorption.  Firstly, \citet{Calzetti+94} showed that the empirical attenuation curve\footnote{We will follow the convention of speaking of `attenuation curves' when the light received on the earth includes light scattered into the beam, and `extinction curves' when the scattered light included in the mean is negligible. \citet{Calzetti+94} considered various source and dust geometries for starburst galaxies (see their sections 4 \& 5) and showed that the geometry of the dust distribution alone could not explain their observational results if the attenuation curve was like the Milky Way or Large Magellanic Cloud and that fundamentally different dust properties were needed.  Their empirical attenuation curve (see their Section 6.2) was derived assuming that the dust is a uniform screen in front of the light source.  This geometry is appropriate for AGNs since the accretion disc and torus restrict our view to the nearside of the AGN.} for star-burst galaxies is flatter than the standard Milky Way curve in the ultraviolet and completely lacks the $\lambda$2175 feature.  \citet{Gaskell+04} then discovered that the mean attenuation curve for AGNs is even flatter in the ultraviolet and also lacking the $\lambda$2175 feature.  Shallow mean attenuation curves for AGNs were also empirically obtained by \citet{Czerny+04} and \citet{Gaskell+Benker07}.  All these curves are shown in Figure 1.  It can be seen that the Calzetti et al. curve is close to the mean AGN curves of \citet{Czerny+04} and \citet{Gaskell+Benker07} except at the very shortest wavelengths.{ The agreement of seven different line and continuum reddening indicators for the well-studied AGN NGC~5548 \citep{Gaskell+23} favors a flatter reddening curve similar to that of \citet{Gaskell+Benker07} for NGC~5548.}

\begin{figure}
 \centering \includegraphics[width=8.3cm]{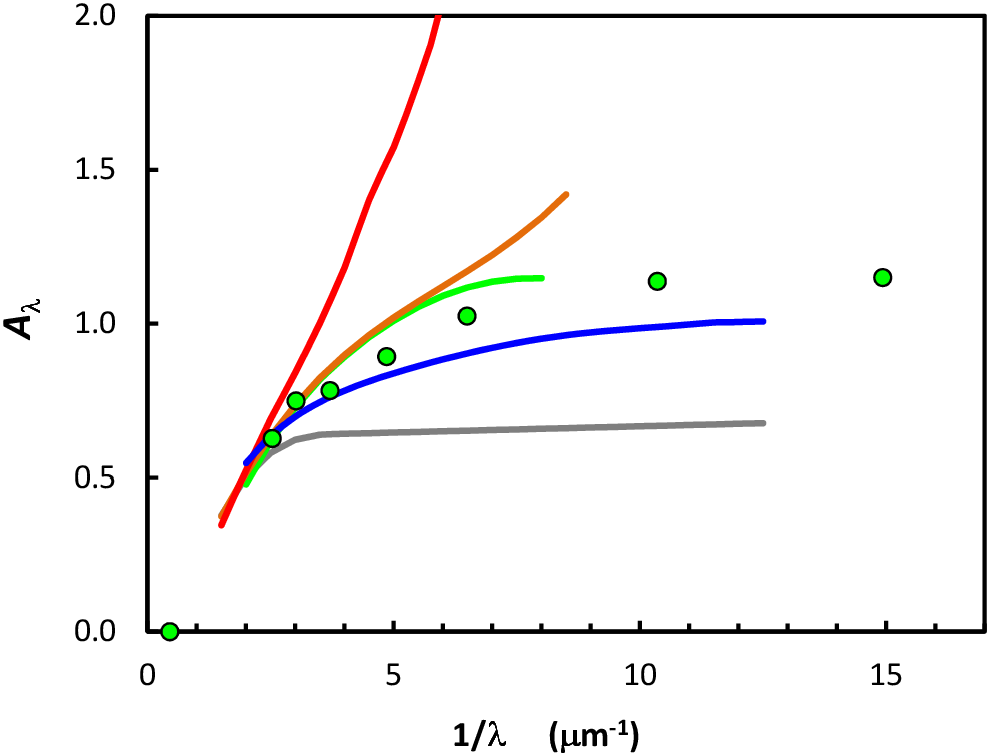}
 \caption{Attenuation estimates (green dots) for non-BALQSOs in the redshift range $1.2 < z < 1.4$ compared with different attenuation and extinction curves.  All attenuations have been set to zero in the rest-frame $K$ band (2.2 $\mu$m) and the curves have been normalized to the observed $V$-band attenuation. From bottom to top the curves are the \citet{Gaskell+04} curve for radio-loud AGNs (grey), the mean curve derived here for non-BALQSOs with redshift $> 1.2$ (blue), the \citet{Gaskell+Benker07} curve (green) for mostly radio-quiet AGNs, the \citet{Calzetti+94} curve for { non-AGN} starburst galaxies (orange), and the SMC extinction curve (red).  The \citet{Czerny+04} curve for optically-selected AGNs is very close to the \citet{Gaskell+Benker07} mean AGN curve and is not plotted separately.}
\end{figure}

An important point \citet{Yamoto+Vansevicius99} showed (see their Figure 4) is that the flatter Calzetti et al.\@ curve can explain the excess reddening of BALQSOs just as well as a steep SMC-like curve. Since the \citet{Czerny+04} and \citet{Gaskell+Benker07} curves are essentially indistinguishable from the Calzetti et al.\@ curve except at the shortest wavelengths (see Figure 1), this means that those curves might explain the reddening of ultraviolet spectral energy distributions (SEDs) of BALQSOs too.  However, for a given change of UV spectral slope, the corresponding $E(B-V)$ for attenuation curves such as the \citet{Gaskell+Benker07} mean AGN curve is greater than for SMC-like curves. To know the total attenuation and to correct the overall SED of BALQSOs it is therefore necessary to know the form of the attenuation curve.

Determining this and the amount of dust associated with the high-velocity outflows of BALQSOs is also important because, as \citet{Scoville+Norman95} pointed out (see also \citealt{Zhang+14}), dusty gas will be accelerated radially outwards by the high radiation pressure efficiency of dust mixed with gas if the gas is exposed to the central radiation field, which is certainly the case for BAL gas.

It has been suggested that the continuum shape of an AGN is a factor in whether an AGN has BALs or not (e.g., \citealt{Baskin+13}).  To ascertain the extent to which this is true we need to know the attenuation and the form of the attenuation curve.

In this paper we therefore determine the attenuation curve for the extra dust associated with BALQSOs.  For brevity we will simply refer to this as `BAL dust'. Strictly speaking we can only say that the dust is along our line of sight and not that is is necessarily mixed in with the BAL gas itself. It could be some other extra extinction that happens to be present on average when BALs are seen, but, as will be discussed below, this seems a less likely explanation.  To determine the mean attenuation curve for BAL dust it is necessary to extend comparisons of observed SEDs to the rest-frame IR. \citet{Maddox+Hewett08} used UKIRT Infrared Deep Sky Survey $K$-band and SDSS $i$-band photometry to investigate the attenuation curves of BALQSOs. They concluded that the attenuation curve was intermediate between an SMC-like curve and the \citet{Gaskell+Benker07} curve.  There are three things to note however: (1) the range of wavelengths used by \citet{Maddox+Hewett08} was only a factor of two, (2) it did not extend into the rest-frame IR of the AGNs, and (3) it covered different spectral regions at different redshifts because it was at fixed observed wavelengths.  In our study, to get around these problems, we use mid-IR photometry from the {\it WISE} satellite in order to obtain the {\em rest frame} 2.2 $\mu$m fluxes for the AGNs.

\section{Sample and Analysis}

We used the large Data Release 10 (DR 10) sample of  166,583 AGNs \citep{Paris+14}.  Since previous work (see above) shows that BALQSOs are more reddened that non-BALQSOs, and hence fainter in the UV, we matched the two samples in the rest-frame $K$ band (2.2 $\mu$m) where the difference in extinction is small.  2.2 $\mu$m is also near the peak of emission by the hottest dust in an AGN and at this wavelength this emission dominates over the host-galaxy starlight.  We estimated the 2.2 $\mu$m rest frame flux by interpolating between the mid-IR photometric points observed by the {\it WISE} satellite.  We excluded AGNs with no {\it WISE} detections or with warning flags.  This reduced the sample to 85,387 AGNs. The BALQSO sample consisted of the subset of 11,211 of these AGNs for which \citet{Paris+14} had assigned a BAL flag.  To extend the reddening curves to shorter wavelengths we used {\it GALEX} $FUV$ ($\lambda$1540) and $NUV$ ($\lambda$2220) photometry.

To eliminate the possibility of a luminosity-dependence of the intrinsic continuum shape we constructed a non-BALQSO sample matched in redshift and 2.2 $\mu$m rest frame luminosity. The list of 85,387 AGNs was sorted by redshift and then for each of the 11,211 BALQSOs we chose two non-BALQSOs of similar redshift and similar 2.2 $\mu$m luminosity.

\subsection{Determining attenuation curves for BAL dust}

Attenuation curves for BAL dust were obtained by comparing the BALQSOs with the luminosity- and redshift-matched, non-BALQSO sample. We determined flux ratios with respect to the rest-frame 2.2 $\mu$m flux for the $FUV$, $NUV$, $u$, $g$, $r$, $i$, and $z$ pass bands {for} each AGN.  AGNs were grouped into redshift bins of (0.4 -- 0.6), (0.6 -- 0.8), (0.8 -- 1.0), . . . , (2.4 -- 2.6).  The reason for matching in redshift and restricting the comparisons to redshift bins was so that the effects of emission lines on the filters were similar for the BALQSOs and non-BALQSOs.  For each redshift bin attenuation curves were determined using the standard pair method by comparing the seven flux ratios. We made no allowance for the small effect of the broad absorption lines themselves on the fluxes in the passbands.

A significant fraction of the AGNs did not have {\it GALEX} detections, especially in the $FUV$ band.  To allow for non-detections, and to minimize the effects of flux measurements with large uncertainties, we used non-parametric statistics.  Although the medians could be compared for most subsamples, we chose to compare upper quartiles because the number of non-detections in the $FUV$ band was $\sim 50$\% for some subsamples.

\subsection{Determining mean attenuation curves for non-BALQSOs}

As a check, and for comparison, we also determined reddening curves by the same method (i.e., comparing 75th percentiles in redshift bins) for the {\em non}-BALQSOs.  We simply divided the {entire sample of} 74,176 {\em non}-BALQSOs into two halves based on the rest frame 2.2 $\mu$m to $\lambda$3000 colour.  We chose $\lambda$3000 because it is a short enough wavelength to reduce the influence of the host galaxy starlight but still at a long enough wavelength that the reddening curves do not deviate too much (see Figure 1).\footnote{Choosing a shorter wavelength would bias the red half of the distribution towards AGNs with steep reddening curves.}

\section{Results}

\subsection{Reddening of non-BALQSOs}

In Figure 1 we show the average attenuation curve for all the non-BALQSOs with $1.2 < z < 2.4$ compared with the various previously-considered reddening curves discussed above. We also show an example of the attenuations in the seven filters for one of the redshift subsamples of non-BALQSOs (the green circles).  Individual AGNs show a variety of attenuation curves \citep{Gaskell+Benker07} but the average curve is relatively flat.  Attenuation curves for non-BALQSOs, the dependence of these curves on various factors, and issues such as the fraction of AGNs with steep SMC-like curves and the effect of selection effects, will be discussed at length in a separate paper. The reason for showing the average attenuation curve for the non-BALQSOs here is to show that our method of analysis described above produces relatively flat AGN attenuation curves for the non-BALQSOs in agreement with previous results \citep{Gaskell+04,Czerny+04,Gaskell+Benker07}.

Because we divided the sample of non-BALQSOs in half by colour, the amount of attenuation we measure is approximately the mean attenuation of { SDSS} non-BALQSOs.  It can be seen that the attenuations shown in Figure 1 correspond to an extinction in the $V$ band of $A_V \thickapprox 0.4$ which corresponds to $E(B-V) \thickapprox 0.15$. { This is similar to the median reddening of SDSS AGNs determined from hydrogen line ratios (see \citealt{Dong+08} and \citealt{Gaskell17}) and to the median reddening \citet{Heard+Gaskell23} found from continuum colours of 4166 SDSS AGNs. It must be remembered that most SDSS AGNs are colour-selected.  This strongly biases them to be blue and hence to have lower reddenings than type-1 thermal  AGNs selected by other means.  This is supported by the higher reddenings found using multiple reddening indicator for well-studied nearby AGNs such as NGC~5548, where the average of seven indicators gives $E(B-V) = 0.26 \pm 0.03$ \citep{Gaskell+23}, and NGC~7469, where the average of six indicators gives $E(B-V) = 0.46 \pm 0.03$ \citep{Kwan+23}.}

\subsection{The attenuation curve for BAL dust in HiBALs}  

\begin{figure}[t!]
 \centering \includegraphics[width=8.3cm]{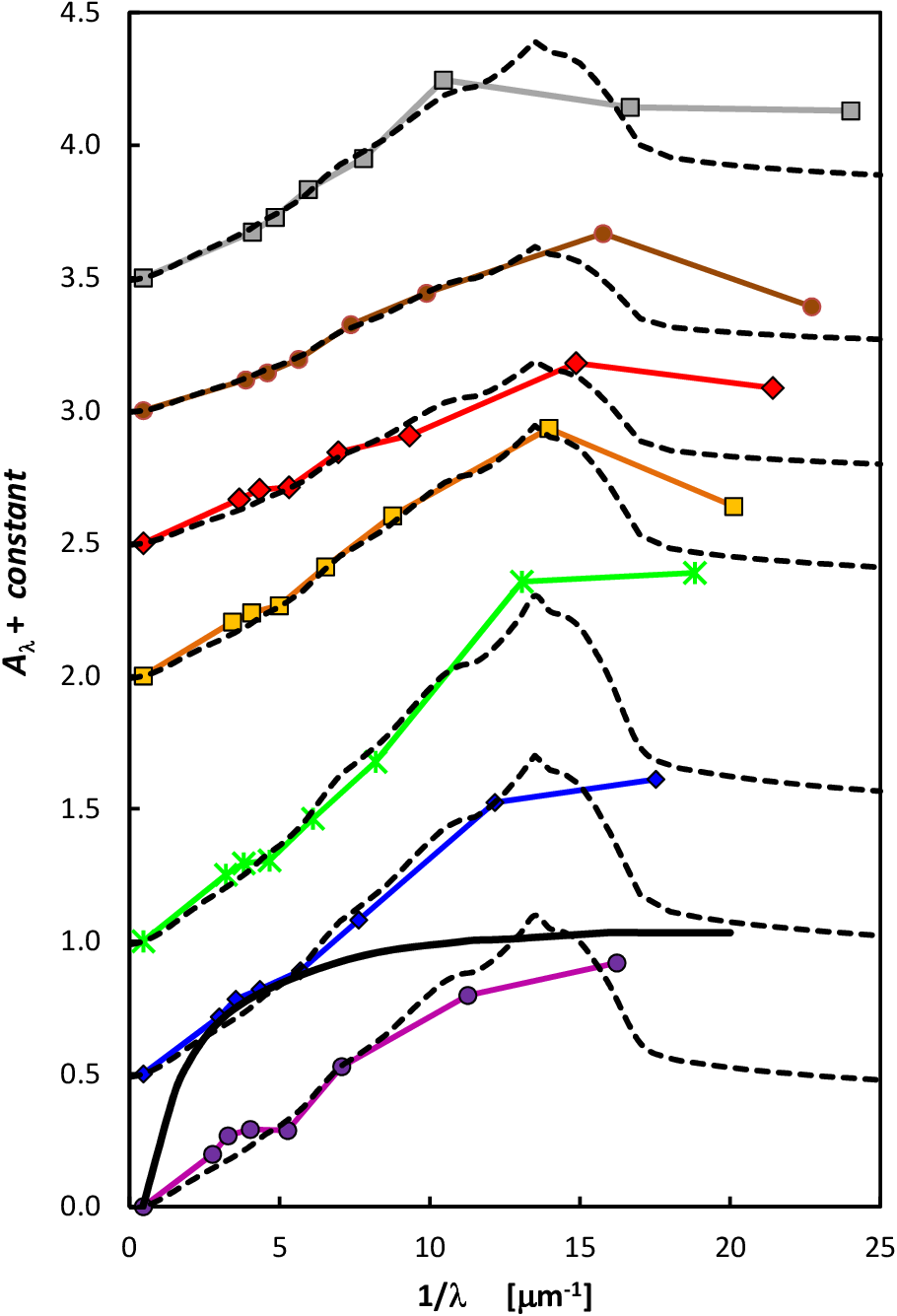}
 \caption{Attenuation curves for the {\em additional} attenuation due to `BAL dust' in HiBALs.  Theoretical SMC curves from \citet{Weingartner+Draine01} (dashed lines) have been fit to the attenuation estimates {for $1/\lambda < 10~\mu$m$^{-1}$} for each redshift bin.  Attenuation curves and fits for successive redshift bins have been displaced upwards by 0.5 or 1.0 magnitudes for convenience. The lowest curves are for $1.4 \leqslant z < 1.6$ and $z$ increases by 0.2 for successive curves.  The average attenuation curve for non-BALQSOs estimated by the same method {(see section 2.2)} is shown as a solid black line at the bottom.  It can be seen that the SMC curve is an excellent fit to each curve for $1/\lambda < 10~\mu$m$^{-1}$. }
\end{figure}

Figure 2 shows the attenuation curves found for { the additional dust associated with the BALs} in {the 16229 HiBAL} QSOs with $z > 1.4$. \citet{Trump+06} find that only $\sim 1$\% of SDSS AGNs are LoBALs so the $z > 1.4$ BALQSOs are overwhelmingly HiBALs.  {For the portions of the curves derived using SDSS photometry, the relative uncertainty in $A_{\lambda}$ at each wavelength is mostly less than $\pm 0.05$ mag. (comparable to the size of the plotting symbols). For the portions using the considerably less accurate {\it GALEX} photometry (the last two points in each curve), the uncertainty is larger and of the order of $\pm 0.2$ magnitudes} 

It can be seen, in striking contrast to the attenuation curves for non-BALQSOs shown in Figure 1 {and to the similar curve derived here for the 74,176 non-BALQSOs\footnote{{Comparison of the non-BALQSO attenuation curve derived here with the previously published curves shown in Figure 1 shows that the non-BALQSO attenuation curve lies between the curves of \citet{Gaskell+04} and \citet{Gaskell+Benker07}.} } (the black curve towards the bottom of Figure 2 -- see Section 2.2)}, that the attenuation curves for the { extra dust in} BALQSOs are all much steeper.  Furthermore, in every case, they are in excellent agreement with the SMC curve for $1/\lambda < 10~\mu$m$^{-1}$ although the mean reddening differs from bin to bin.\footnote{It should be remembered that SDSS AGNs are mostly colour selected and that the colour selection is redshift dependent.}  There is no evidence for redshift or luminosity dependence in the {\em shape} of the attenuation curves. It can also be seen that, as expected, the broad absorption lines themselves in the spectra have only a small effect on the fluxes in a passband.  For example, broad \ion{C}{iv} and \ion{Si}{iv} absorption troughs will cause the exinction at 7 $\mu$m$^{-1}$ to be a little higher.  This is barely detectable in Figure 2.

\subsection{The attenuation curve for BAL dust in LoBALs} 

\begin{figure}
 \centering \includegraphics[width=8.3cm]{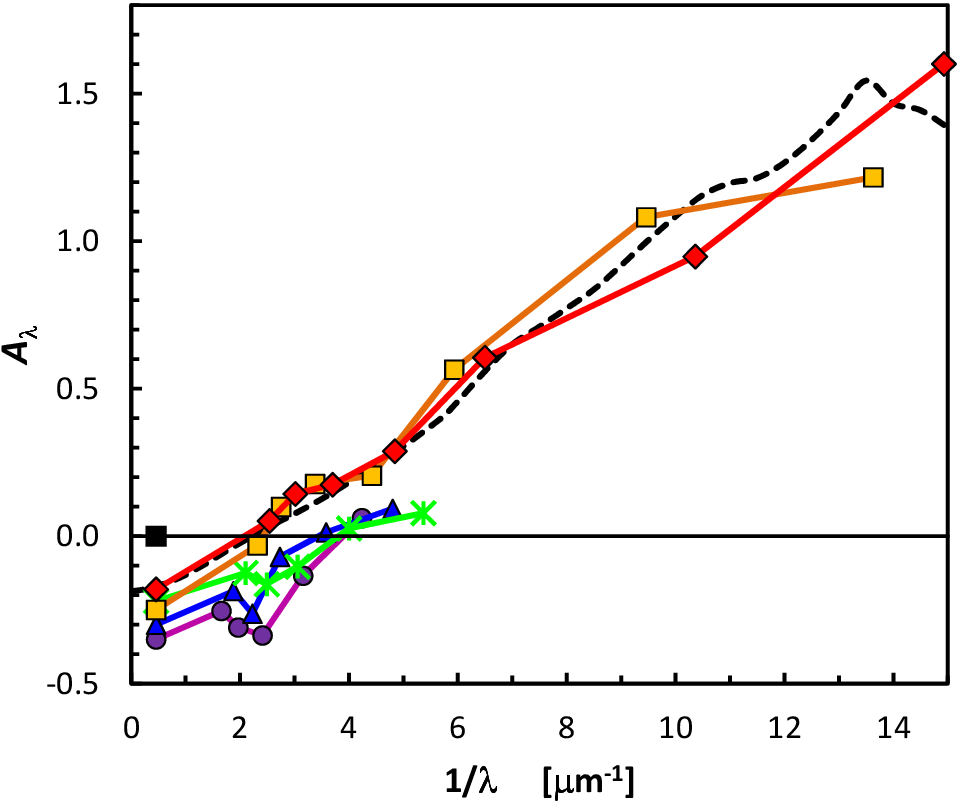}
 \caption{Attenuation curves for the excess extinction in LoBALs in different redshift bins.  The purple lines and circles are for $0.4 \leqslant z < 0.6$, the blue lines and triangles for $0.6 \leqslant z < 0.8$, the green lines and asterisks for $0.8 \leqslant z < 1.0$, the orange lines and squares for $1.0 \leqslant z < 1.2$, and the red lines and diamonds for $1.2 \leqslant z < 1.4$.  The black dashed line is the theoretical SMC curve of \citet{Weingartner+Draine01} normalized to $E(B-V) = 0.05$ to match the observed attenuations for the two highest redshift bins.  By construction, all of the attenuation curves would pass through zero at 2.2 $\mu$m (shown by the black square) but an additional, lower symbol has been added on the left at 2.2 $\mu$m for each attenuation curve to indicate the offset due to added scattered light -- see discussion in the text. (The attenuations in the {\it GALEX} bands have been omitted for the lower-redshift subgroups because the smaller sample sizes in these bins coupled with the large uncertainties in the {\it GALEX} fluxes makes the comparison very uncertain.)}
\end{figure}

Figure 3 shows the attenuation curves for { the additional dust associated with} {the 232} BALQSOs with $z < 1.4$. These AGNs are all LoBALs since broad \ion{C}{iv} absorption cannot be detected in SDSS spectra for $z < 1.4$. An unexpected result is that the { additional} attenuation of these LoBALs is {\em negative} in the optical and shows a downward slope from the rest-frame $K$ band (the black square in Figure 3).  Negative attenuations means that there is either extra light at shorter wavelengths in the LoBALs compared with the non-BALQSOs or less emission at 2.2 $\mu$m (see discussion in Section 4.3 below). { Other than this difference, it can be seen from Figure 3 that the attenuation curve of the additional dust in Lo-BALs is similar to the the curves of the additional dust HiBALs in Figure {2} and to an SMC-like extinction curve.}

\subsection{Magnitude of the extinction by BAL dust}

From each fit of an SMC curve to an observed attenuation curve we obtain $E(B-V)$. We show the distribution of  $E(B-V)$ for the LoBALs and HiBALs in Figure 4.
The mean $E(B-V)$ for the HiBALs is 0.027, in good agreement with previous studies using SDSS-selected BALQSOs (e.g., \citealt{Reichard+03,Dai+08,Pu14}).  This is several times less than the optical reddening for a typical AGN from normal AGN dust, as can be seen by comparing the attenuation curve for the lowest redshift sample in Figure 2 with the mean reddening curve for non-BALs.  The additional attenuation in the optical for HiBALs is thus of the order of $\sim 20$\% of the total optical extinction. For the $z < 1.4$ LoBALs the mean $E(B-V)$ is 0.045. In agreement with previous studies (e.g., \citealt{Reichard+03,Pu14}), this is greater than for the HiBALs.  However, we note that the HiBALs show a decline in $E(B-V)$ with redshift and that the LoBALs are consistent with the low-redshift extrapolation of this (see Figure 4).  Since we have a magnitude-limited sample, the decline with redshift corresponds to a decline in $E(B-V)$ with increasing luminosity.  This translates into a mean luminosity increase by a factor of 5 for the LoBALs as the redshift increases, and a factor of 3 for the HiBALs.  It will clearly be important in future studies to investigate the possibility of luminosity-dependent reddening of BALQSOs and the effect this could have on detection rates for BALQSOs of all classes (see, for example, \citealt{Dai+12}).

\begin{figure}
 \centering \includegraphics[width=8.3cm]{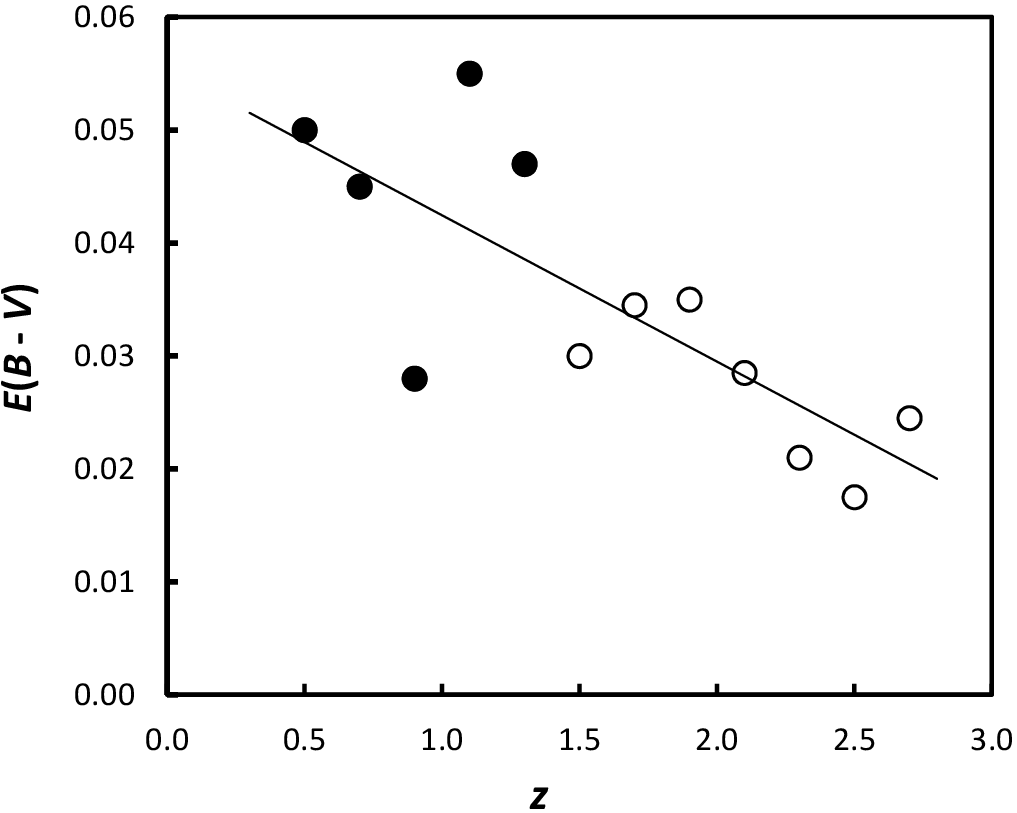}
 \caption{The mean reddenings, $E(B-V)$ for each redshift bin.  Filled circles are LoBALs and open circles are HiBALs.  {Uncertainties in the means are $\pm 0.005$ magnitude for the HiBALs and $\pm 0.012$ magnitude for the LoBALs.} The line is a least-squares fit {to both samples}.}
\end{figure}

\section{Discussion}

\subsection{The attenuation curve for BAL dust}

We have combined the attenuation curves shown in Figure 2 to produce a mean attenuation curve for BAL dust.  This is shown in Figure 5.  This clearly shows that BAL dust does indeed have an SMC-like attenuation curve, as has generally been assumed.  This is true both for LoBALs and HiBALs. As can be {seen} from Figures 2 and 5, there is close agreement between the BAL dust attenuation curves and the SMC extinction curve for $1/\lambda < 10~\mu$m$^{-1}$.

\subsection{Extending the SMC-type extinction curve to the extreme ultraviolet}

Because the extinction curve for the SMC has been determined from comparison of fluxes of stars at low redshift, it has only been determined observationally to $1/\lambda = 8~\mu$m$^{-1}$ \citep{Rocca-Volmerange+81,Nandy+82,Prevot+84}.  To extrapolate to $1/\lambda >~8~\mu$m$^{-1}$ one has to depend on theoretical extinction curves from grain models.  In Figures 2 and 3 we have used the theoretical SMC extinction curve model of \citet{Weingartner+Draine01}. Our composite BAL attenuation curve (Figure 5) provides important support for their prediction that an SMC-like curve continues to rise steeply out to $1/\lambda \geq  14~\mu$m$^{-1}$. For $1/\lambda = 14~\mu$m$^{-1}$ the Weingartner \& Draine SMC model predicts a sharp drop in the extinction.  Although the uncertainties are larger for attenuations calculated from the {\it GALEX} filters, our results (see Figure 5) show that the attenuation curve does indeed continue its sharp rise into the extreme ultraviolet (EUV) and then declines.  However, if the SEDs of BALQSOs and non-BALQSOs are intrinsically the same in the EUV, as we are assuming, the decline is not as great as predicted by the Weingartner \& Draine SMC model.  

{ It has long been known that the relative 2 keV X-ray flux, as given by the optical to X-ray spectral index ($\upalpha_{ox}$), is lower for AGNs with absorption lines than for those without \citep{Brandt+00}. For BALQSOs the deficit in 2 keV X-rays is an order of magnitude, corresponding to $\upalpha_{ox}$ being steeper by $\Delta \upalpha_{ox} \thickapprox 0.4$.  This remains true when {one} allows for the systematic steepening of $\upalpha_{ox}$ with increasing optical/UV luminosity (see Figure 3 of \citealt{Liu+18})\footnote{The luminosity dependence of $\upalpha_{ox}$ should not {be} a factor when BALQSOs and non-BALQSOs are matched in IR luminosity as we have done here.}. The cause of the observed weaker relative X-ray fluxes in BALQSOs is not clear.  The first proposal, made by \citet{Brandt+00}, was that the remarkable correlation of relative X-ray weakness with increased absorption pointed to this absorption itself as the primary cause of soft X-ray weakness in QSOs.  Support for this idea comes from deep {\it XMM-Newton} observations of the $z = 6.5$  X-ray weak BALQSO J0439+1634.  \citet{Yang+22} find an intrinsic column density $N_H \thickapprox 2.2 - 6.1 \times 10^{23}$ cm$^{-2}$, which could lower the X-ray flux by an order of magnitude.  If the hypothesis of \citet{Brandt+00} is correct and the differences in X-ray properties are primarily due to absorption, the observed X-ray differences are not a problem for our reddening curve derivation in the EUV.}

{ Another possible explanation of the observed X-ray weakness is that BALQSOs have, on average, an {\em intrinsically} steeper $\upalpha_{ox}$ and that this somehow favours the launching of BAL outflows.  It is well known that BALQSOs have high Eddington ratios \citep{Boroson+Meyers92,Yuan+Wills04}.  However, \citet{Vito+18} found no evidence for an Eddington-ratio dependence of X-ray weakness in BALQSOs.}  

{ Given the observed X-ray weakness of BALQSOs we cannot at present exclude the possibility that the EUV spectrum of BALQSOs below 700 \AA~might be intrinsically weaker than for non-BALQSOs.  If this is the case, it would lead to an overestimate of the reddening and could thus explain the slight discrepancy at the highest energies between the attenuation we find and the theoretical SMC reddening curve of \cite{Weingartner+Draine01}} It would be interesting in the future to construct EUV attenuation curves for samples matched in the equivalent width of the \ion{He}{ii} $\lambda$1640 emission line (see \citealt{Baskin+13}) since this is a good probe of the intrinsic EUV continuum.

If the difference between the observed and predicted curves beyond 15 $\mu^{-1}$ is real, it could imply a somewhat different grain size distribution from that assumed in the Weingartner \& Draine SMC model (B. T. Draine, private communication).

\begin{figure}
 \centering \includegraphics[width=8.3cm]{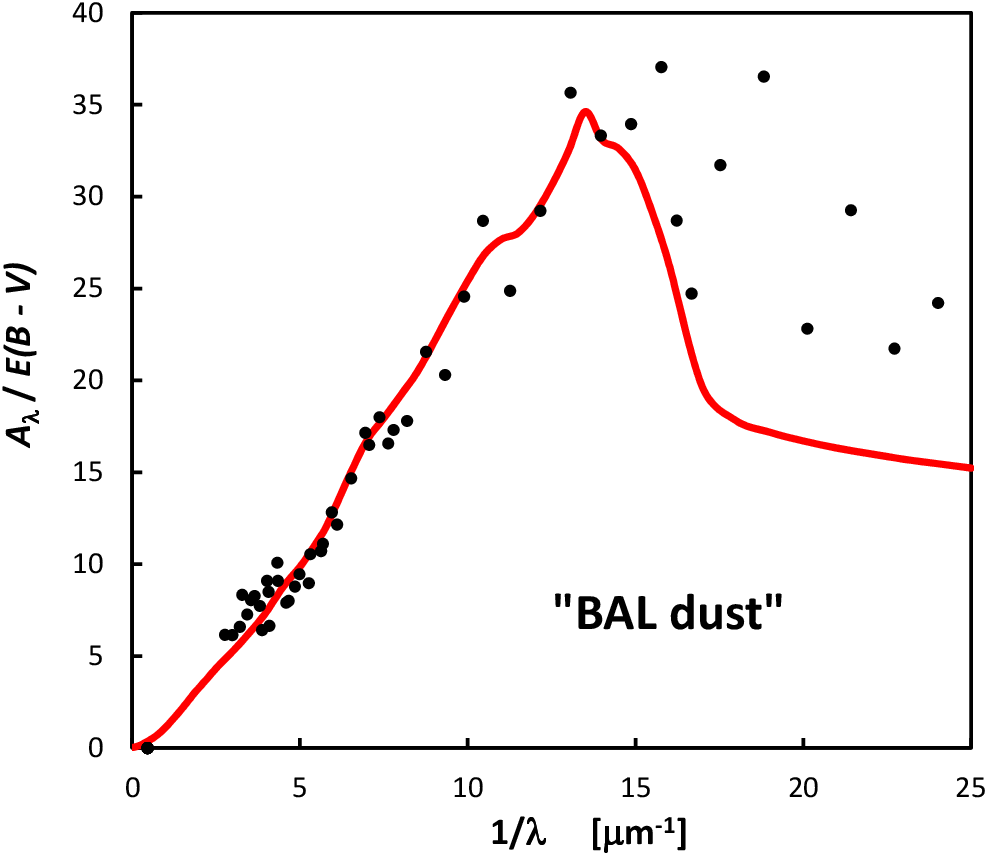}
 \caption{Attenuations for BALQSOs (points) compared with the theoretical SMC extinction curve (red curve) of \citet{Weingartner+Draine01}.  The points are taken from the redshift subsets shown in Figure 2, but they have been divided by the $E(B-V)$ of the subset determined from the fits to the theoretical SMC curves {shown in Figure 2.  These fits were determined} for  $1/\lambda < 10~\mu$m$^{-1}$.}
\end{figure}

\subsection{The origin of the `negative' attenuation in LoBALs}

Perhaps our most surprising result (see section 3.3 and Figure 3) is the apparently {\em negative} attenuation for LoBALs.  This implies that, compared with non-BALQSOs, the LoBALs have extra light relative to their rest-frame 2.2 $\mu$m fluxes on average.  This could either be because the LoBALs have less reprocessing of radiation by dust (i.e., by the hot dust dominating the 2.2 $\mu$m emission) or that they have additional light at shorter wavelengths.  Given that the LoBALs have higher extinction and hence more dust in general, having less thermal emission by dust would seem unlikely.  Instead, the easiest explanation is that, on average, {\em LoBALs have more scattered light} than non-BALQSOs and HiBALs.  In this scenario there is more dust causing extinction along the line of sight in LoBALs, as is observed, but also more dust off the line of sight causing scattered light.  As can be see in Figure 3, a simple decrease of the $K$-band fluxes by $\sim 25$\% or a corresponding addition of scattered light at shorter wavelengths brings the shape of the attenuation curves into agreement with the SMC-like curve like the HiBALs.  Since dust is strongly forward scattering, a large covering factor is not needed so long as the dust doing the scattering is close to our line of sight.  We are assuming here that the scattered light spectrum is relatively flat. This will be the case if the scattering dust is optically thick (see \citealt{Zubko+Laor00,Kishimoto+01} and discussion in section 4.4 of \citealt{Goosmann+Gaskell07}).  

\subsection{Explaining apparently `steeper-than-SMC' extinction curves}

After correction for the slight difference in relative mean rest-frame 2.2 $\mu$m fluxes (see section 3.3) the overall attenuation curve for LoBALs is well fit by an SMC curve (see Figure 3). There are, however, some small-scale deviations from the SMC curve.  We believe that these small features are due to real spectral differences between the LoBALs and the control non-BALQSO sample.  In Figure 3 there is a dip in the extinction curve around 4000 - 5000 \AA. Inspection of the spectra of the LoBALs shows that this is indeed a result of a difference in the spectra, the well-known, strong, broad optical \ion{Fe}{ii} emission in LoBALs \citep{Weymann+91}.

The higher-redshift LoBALs in Figure 3 show a local flattening of the attenuation curve from $\lambda$3000 to $\lambda$2000.  We suspect this to be a result of the different \ion{Fe}{ii} emission properties of LoBALs and non-BALQSOs -- in this case the \ion{Fe}{ii} emission of the so-called `small blue bump'.   \citet{Hall+02} have suggested that the UV SEDs ($\lambda < 3000$ \AA) of some LoBALs require a reddening curve that is steeper than the SMC curve.  We believe this to be due in some cases to the apparent (but probably spurious) flattening of our LoBAL attenuation curve from $\lambda$3000 to $\lambda$2000  just discussed.  It has to be recognized that when there is no observational coverage to the IR, one has freedom in how to normalize an observed attenuation curve.  \cite{Hall+02} did not have observations longwards of $\lambda$3000.  If one chooses to normalize an SMC curve from $\lambda$3000 to $\lambda$2000, then attenuation shortwards of $\lambda$2000 appears to rise faster than an SMC curve.  However, {if} one normalizes an SMC curve {\em shortwards} of $\lambda$2000 this gives an attenuation curve from $\lambda$3000 to $\lambda$2000 that is flatter than an SMC curve.  In our case we have longer-wavelength observations (see Figure 3) and these show that the attenuation curve is just locally flat from $\lambda$3000 to $\lambda$2000.

Although a steep SMC-like reddening curve usually successfully makes the SED of an non-BALQSO match a BALQSO (see, for example, the spectra in \citealt{Urrutia+09}), there are AGNs for which observations to longer wavelengths suggest that the attenuation curve in some cases continues to be flat to long wavelengths \citep{Meusinger+05,Zhou+06,Fynbo+13,Jiang+13}).  For the attenuation curves for the $1.0 < z < 1.4$ AGNs in Figure 3, note that if there were no information for wavelengths longer than 5000\AA, the curves would appear to be rising faster than an SMC curve.

We suggest that in cases where there is a flatter attenuation curve at long wavelengths another explanation could be partial coverage of the inner regions of the AGN by the BAL dust.  There have long been multiple lines of evidence for partial coverage by BAL gas (see, for example, \citealt{Hamann+02}, \citealt{Gaskell+Harrington18} and references therein).  This means that the size of a typical BAL cloud is similar to, or smaller than, the size of the background continuum light source.  If we are correct that the BAL dust is mixed with the BAL gas, there will be partial coverage of the continuum by the dust.  BALQSOs are selected by their UV absorption lines, so the BAL clouds partially or totally cover the part of the accretion disc emitting the UV.  If the BAL cloud is comparable in size to this inner region of the disc, then the cloud will quite likely be smaller than the much larger outer region dominating the optical emission.  We find that construction of a simple toy model with this geometry easily produced an effective attenuation curve that was SMC-like at shorter wavelengths and flat at longer wavelengths.

\subsection{The nature of LoBALs}

A popular idea for the origin of LoBALs (and especially the extreme FeLoBALs) is that they are `young' AGNs.  In such models, the AGN is initially obscured inside a dusty starburst until what has been called the `blowout' phase when a strong outflow driven by the starburst or by an outflow from the AGN itself stops the starburst and uncovers the AGN (see, for example, \citealt{Hopkins+05}).

The SMC-like attenuation curve we find for the extra attenuation in HiBALs and LoBALs does not favour LoBALs being young AGNs breaking out of their cocoons.  This is because, both in the Milky Way and in starburst galaxies, star-forming regions do {\em not} have SMC-like extinction curves \citet{Calzetti+94}.  This difference, plus the similarity of HiBAL attenuation curves to LoBAL attenuation curves, suggests that {\it the dust is associated with outflows themselves} and not with a star-forming cocoon around a young AGN.

As \citet{Herbst+16} note, despite the popularity in the AGN feedback community of the idea of LoBALs being young AGNs, there is little observational support for this picture.  \citet{Violino+16} find no evidence from sub-mm observations of FeLoBALs for cold starbursts and they find that the sub-mm properties of matched samples of BALs and non-BALs are indistiguishable.  From {\it HST} imaging \citet{Herbst+16} find that FeLoBALs appear to reside in faint, compact hosts with weak or absent merger signatures.

\subsection{Acceleration of BAL outflows}

High-velocity outflows of early-type stars result from the absorption and scattering of line radiation from the intense radiation field \citep{Castor+75,Cassinelli79}.  A similar mechanism has been widely considered to drive outflows in BALQSOs (e.g., \citealt{Drew+Boksenberg84}).  The effectiveness of radiation pressure in driving winds depends on the fraction of energy (and hence momentum) removed from the radiation field.  For line-driven winds it depends on the fraction of the spectrum intercepted by the blueshifted resonance lines.  If the absorbing gas is too highly ionized (`overionized'), there are fewer resonance lines to intercept the radiation and the acceleration is less effective.  However, if there is dust associated with the gas this also absorbs radiation.  If the dust and gas are coupled, as will happen if the grains are charged, the radiation pressure on the dust will drive the outflow (see \citealt{Cassinelli79}).  Such dust-driven winds are well known and drive the very high mass loss rates in cool stars.

The relative effectiveness of line-driving versus dust-driving is easy to derive empirically -- one simply needs to know how much energy is taken out of the spectrum by each mechanism. { Figure 6 shows two independent estimates of the spectral energy distribution (SED) of AGNs and the effect of SMC dust with $E(B-V) = 0.04$. One of the AGN SEDs is the de-reddened SED of NGC~5548, an empirically-constructed SED derived by \citet{Gaskell+07}(GKN) from observations. The other is the widely-used SED of \citet{Mathews+Ferland87} derived primarily from photoionization and energy-balance considerations from many AGNs.\footnote{This is the continuum implemented as TABLE AGN in the photoionization code {\it Cloudy} \citep{Ferland+98,Ferland+17}.}  As can be seen, the two SEDs agree well from the optical to hard X-ray regions.\footnote{The 
 factor of $\sim 2$ greater IR flux of \citet{Mathews+Ferland87} is because they derived the relative IR emission as seen from the earth (i.e., outside the IR-emitting regions),
whilst \citet{Gaskell+07} derived the IR emission as seen by the BLR, which is located {\em inside} the dust-emission region.} Figure 6 shows how the far-UV continuum we see is attenuated by the dust in a typical BAL outflow.  For a reddening curve like the theoretical \citet{Weingartner+Draine01} SMC curve, about half of the energy of the Big Blue Bump is removed.  This will be especially true if the attenuation at the highest energies (shortwards of 10 $\upmu$m$^{-1}$) is larger than calculated by  \citet{Weingartner+Draine01}, as our reddening estimates indicate (see Figure 5 and discussion in Section 4.2). One can thus readily appreciate} that far more energy is taken out of the spectrum by dust than would be taken out by absorption lines,  { since even a very broad absorption line covers a small range in frequency.  For example, an extremely broad absorption line with a width of 30,000 km s$^{-1}$ only blocks the SED over a range of 0.04 dex in $\log \nu$.} Although the attenuation { due to dust} is low in the optical, it is of the order of a whole magnitude in the EUV, and, { as can be seen in Figure 6,} this is precisely where the energy output of AGNs is greatest.

\begin{figure}
 \centering \includegraphics[width=8.3cm]{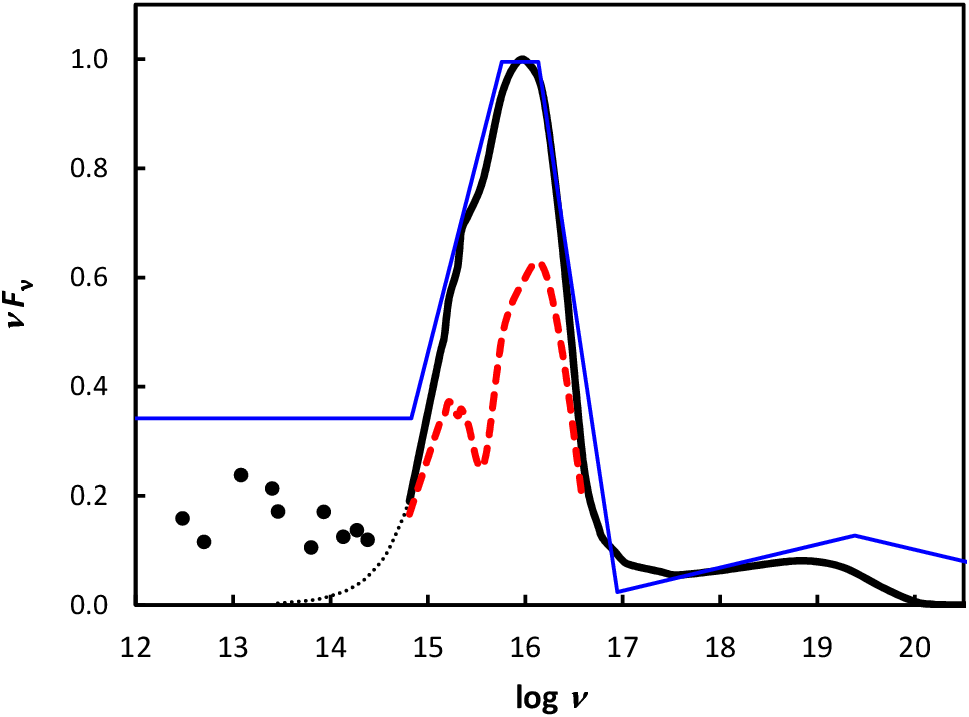}
    \caption{ The amount of energy removed by dust in a typical BAL outflow.  In a $nu F_{\nu}$ versus $\log \nu$ plot such as this, the area under the curves is power.  The solid black curve and black dots show the de-reddened spectral energy distribution (SED) of NGC 5548 empirically constructed by \citet{Gaskell+07} from observations.  For comparison, the thin blue line shows the \citet{Mathews+Ferland87} AGN continuum derived primarily from photoionization and energy-balance considerations.  The dashed red curve shows the GKN continuum reddened by the \citet{Weingartner+Draine01} SMC curve for $E(B-V) = 0.05$. (Plot adapted from \citealt{Gaskell08}).}
\end{figure}

{ Figure 6 strongly supports what was} suggested by \citet{Scoville+Norman95} and discussed more recently by \citet{Zhang+14}, that it is {\em radiation pressure on dust} which drives BAL outflows.  Furthermore, one magnitude of extinction, corresponding to an optical depth, $\tau \sim 1$, is the optimum value for driving an outflow.  This is because if $\tau \ll 1$, little radiation, and hence little momentum, is absorbed, while if $\tau \gg 1$, the acceleration is low because of mass loading.

Strong broad lines of abundant species like \ion{C}{iv} are typically very optically thick.  When \ion{P}{V} is present the column density, $N_H$ can be $> 10^{23}$ cm$^{-2}$ (e.g., \citealt{Moravec+16}). \citet{Hamann+02} point out that when $N_H$ is this large, it presents a serious challenge to models of radiatively-driven BAL outflows.  The large fraction of radiation intercepted by dust removes this problem.

{ Another advantage of radiative acceleration via dust is that it produces a ``volumentric" force (i.e., a force throughout the cloud rather than merely on the surface).  A volumetric force is necessary for cloud stability and radiation-pressure confinement (see \citealt{Blumenthal+Mathews79} for detailed discussion.)}

\subsection{The dust-to-gas ratio in BAL clouds}

While the mean reddening of BAL outflows is fairly well determined here, there is uncertainty in column densities.  If a HiBAL has a column density of $N_H \sim  10^{20} - 10^{21}$ cm$^{-2}$,
this corresponds to $E(B-V) \sim 0.02 - 0.3$ if we assume a standard Galactic gas-to-dust ratio of $N_H/E(B-V) = 5.8 \times 10^{21}$ \citep{Bohlin+78}.  The reddenings found here are thus consistent with a Galactic gas-to-dust ratio.  {
from comparison of a large number of AGN reddening estimates and X-ray column density determinations, \citet{Jaffarian+Gaskell20} find that, on average, $N_H$ is consistent with predictions from $E(B-V)$ for a solar-neighbourhood dust-to-gas ratio.\footnote{{Earlier claims of AGNs having a much lower average gas-to-dust ratio than the solar neighbourhood (see, for example, \citealt{Maiolino+01} and references therein) are a consequence of the earlier studies being biased to objects with high $N_H$. For example, compared with \citet{Jaffarian+Gaskell20}, the reddenings in \citet{Maiolino+01}, are comparable (median $E(B-V)$ values of 0.77 and 0.66 respectively), but the median $\log N_H$ of the objects considered by \citet{Maiolino+01} is 22.53 versus 21.26 for \citet{Jaffarian+Gaskell20}, and hence 20 times greater.}}}

\subsection{The spectral energy distributions of BALQSOs}

Given the success of an SMC-like curve in explaining the differences between BALQSOs (both LoBALs and HiBALs) and non-BALQSOs, we find no evidence for major differences between the intrinsic SEDs of BALQSOs compared with non-BALQSOs { for the range of wavelengths we consider}.  Previously suggested differences (such as an apparent IR excess) { in BALQSOs} are simply the effects of reddening.  The apparent similarity of the underlying SEDs creates problems for the idea that whether or not a BAL system is formed is determined by the SED.  At best, any systematic difference for HiBAL SEDs has to be at wavelengths shorter than 400 \AA~{ (see Section 4.2)}.

\subsection{Comparing emission-line properties of BALQSOs and non-BALQSOs}

The emission-line properties of BALQSOs and non-BALQSOs show some subtle differences.  \citet{Corbin90} found BALQSOs to have lower \ion{C}{iv} emission-line equivalent widths and a greater blueshifting of the \ion{C}{iv} emission-line.  \citet{Weymann+91} suggested that these small emission-line differences could be due to a systematic difference in viewing angle, coupled with a mild anisotropy in the angular distribution of the flux of the emission lines.  We suggest instead that the differences are not real but are a consequence of matching AGNs in the UV.  Because of the greater extinction of BALQSOs, a BALQSO of the same apparent UV brightness as a non-BALQSO is really intrinsically more luminous.  The equivalent width of \ion{C}{iv} emission declines with increasing luminosity (the `Baldwin effect' -- \citealt{Baldwin77,Baldwin+78}), and so will be slighly lower for the BALQSO.  There will be a systematically slightly greater blueshift of \ion{C}{iv} since the blueshifting is correlated with the Baldwin effect \citep{Corbin90}.  These emission-line differences should not show up in samples matched in the rest-frame IR.

\subsection{The diversity of reddening curves seen in AGNs}

\citet{Gaskell+Benker07} present reddening curves for individual AGNs with near-simultaneous UV and optical spectra and find a diversity of reddening curves including one SMC-like reddening curve. Their mean reddening curve is indistinguishable from the \citet{Czerny+04} one.  Looking at red AGNs naturally favors finding steep, SMC-like reddening curves. From a study of red SDSS AGNs \citet{Richards+03} suggested that the reddening curve for AGNs in general is an SMC-like curve. However, if one considers {\it all} AGNs, the mean reddening curve is flat. For example, \citet{Czerny+04}, using the standard pair method, get a relatively flat reddening curve (shown in Figure 1) from SDSS AGNs. What is significant is that they used {\it essentially the same sample of SDSS AGNs} as \citet{Richards+03}. An indication that an SMC-like curve is not appropriate for all AGNs is the statement by \citet{Richards+03} that attempting to explain their large color-segregated composites generally with reddening ``results in good matches at both 1700 and 4040~\AA~but overpredicts the flux between these two wavelengths and underpredicts the flux shortward of \ion{C}{iv}.'' As \citet{Gaskell+04} note, this is exactly what is predicted by the flatter AGN reddening curve. The cause of the diversity of reddening curves of AGNs, and, in particular, why most AGNs show relatively flat reddening curves will be discussed in a separate paper.

\section{Conclusions}

We have derived attenuation curves for the dust causing the extra attenuation of BALQSOs.  Our main conclusions are as follows:

\begin{itemize}

\item Unlike the flat average attenuation curves for non-BALQSOs, we find a steeply rising, SMC-like attenuation curve for both HiBALs and LoBALs.  We have derived this curve to shorter wavelengths than any previous reddening curves.  We find good agreement with the \citet{Weingartner+Draine01} theoretical extrapolation of the SMC curve to $1/\lambda = 14~\mu$m$^{-1}$ (700 \AA~or $\approx 17$ eV ) but our curve shows somewhat higher extinction out to 400 \AA~($\approx 30$ eV).

\item The steeply-rising attenuation curve for the dust associated with LoBALs is quite different from the relatively flat extinction curve for starburst galaxies and the extinction curves for star-forming regions in the Milky Way.  This does not support the idea that LoBALs are young AGNs in the process of breaking out of a cocoon of dust.  Instead, the BALQSO attenuation curves favor the extra dust in both LoBALs and HiBALs being associated with the high-velocity outflow.

\item The differences in the apparent IR to EUV SEDs of BALQSOs and non-BALQSOs are consistent with being due to dust and not to intrinsic differences.

\item These differences in apparent SEDs clearly show that the radiative acceleration of the BAL clouds is dust-driven, like outflows from cool stars, rather than line-driven, like the winds from O stars.

\item LoBALs have their attenuation curves offset in a manner which is consistent them having more scattered light in the optical and UV than HiBALs and non-BALQSOs.  This scattered light and also possible partial coverage of the central regions of the AGN are probable explanations of attenuation curves that appear to be steeper than an SMC curve.

\end{itemize}

\section*{Acknowledgments}

We are grateful to Ski Antonucci, Bruce Draine, TingGui Wang, Wen-Juan Liu, Ari Laor and the late Bill Mathews for useful discussions and comments.  We also thank the referee for a careful reading and for useful comments.  JJMG and JS carried out their work under the auspices of the Science Internship Program (SIP) of the University of California at Santa Cruz.  We wish to express our appreciation to Raja GuhaThakurta for his excellent leadership of the SIP program.

\section*{Data availability}

All photometric data are available online at http://www.sdss3.org/dr10/algorithms/qso\_catalog.php



\begin{thebibliography}{99}

\bibitem[\protect\citeauthoryear{Baldwin}{1977}]{Baldwin77} Baldwin J.~A., 1977, ApJ, 214, 679. doi:10.1086/155294

\bibitem[\protect\citeauthoryear{Baldwin et al.}{1978}]{Baldwin+78} Baldwin, J.~A., Burke, W.~L., Gaskell, C.~M., \& Wampler, E.~J.\ 1978, Nature, 273, 431

\bibitem[\protect\citeauthoryear{Baskin et al.}{2013}]{Baskin+13} Baskin, A., Laor, A., \& Hamann, F.\ 2013, MNRAS, 432, 1525

\bibitem[\protect\citeauthoryear{Blumenthal \& Mathews}{1979}]{Blumenthal+Mathews79} Blumenthal G.~R., Mathews W.~G., 1979, ApJ, 233, 479. doi:10.1086/157408

\bibitem[\protect\citeauthoryear{Bohlin et al.}{1978}]{Bohlin+78}  Bohlin, R.~C., Savage, B.~D., \& Drake, J.~F.\ 1978, ApJ, 224, 132

\bibitem[\protect\citeauthoryear{Boroson \& Meyers}{1992}]{Boroson+Meyers92} Boroson T.~A., Meyers K.~A., 1992, ApJ, 397, 442. doi:10.1086/171800

\bibitem[\protect\citeauthoryear{Brandt, Laor, \& Wills}{2000}]{Brandt+00} Brandt W.~N., Laor A., Wills B.~J., 2000, ApJ, 528, 637. doi:10.1086/308207

\bibitem[\protect\citeauthoryear{Brotherton et al.}{2001}]{Brotherton+01} Brotherton, M.~S., Tran, H.~D., Becker, R.~H., et al.\ 2001, ApJ, 546, 775

\bibitem[\protect\citeauthoryear{Calzetti et al.}{1994}]{Calzetti+94} Calzetti, D., Kinney, A.~L., \& Storchi-Bergmann, T.\ 1994, ApJ, 429, 582

\bibitem[\protect\citeauthoryear{Cassinelli}{1979}]{Cassinelli79} Cassinelli, J.~P.\ 1979, ARA\&A, 17, 275

\bibitem[\protect\citeauthoryear{Castor et al.}{1975}]{Castor+75} Castor, J.~I., Abbott, D.~C., \& Klein, R.~I.\ 1975, ApJ, 195, 157

\bibitem[\protect\citeauthoryear{Corbin}{1990}]{Corbin90} Corbin, M.~R.\ 1990, ApJ, 357, 346

\bibitem[\protect\citeauthoryear{Czerny et al.}{2004}]{Czerny+04} Czerny, B., Li, J., Loska, Z., \& Szczerba, R.\ 2004, MNRAS, 348, L54

\bibitem[\protect\citeauthoryear{Dai et al.}{2008}]{Dai+08} Dai, X., Shankar, F., \& Sivakoff, G.~R.\ 2008, ApJ, 672, 108-114

\bibitem[\protect\citeauthoryear{Dai et al.}{2012}]{Dai+12} Dai, X., Shankar, F., \& Sivakoff, G.~R.\ 2012, ApJ, 757, 180

\bibitem[\protect\citeauthoryear{Drew \& Boksenberg}{1984}]{Drew+Boksenberg84} Drew, J.~E., \& Boksenberg, A.\ 1984, MNRAS, 211, 813

\bibitem[\protect\citeauthoryear{Dong et al.}{2008}]{Dong+08} Dong, X., Wang, T., Wang, J., et al.\ 2008, MNRAS, 383, 581

\bibitem[\protect\citeauthoryear{Ferland et al.}{1998}]{Ferland+98} Ferland G.~J., Korista K.~T., Verner D.~A., Ferguson J.~W., Kingdon J.~B., Verner E.~M., 1998, PASP, 110, 761. doi:10.1086/316190

\bibitem[\protect\citeauthoryear{Ferland et al.}{2017}]{Ferland+17} Ferland G.~J., Chatzikos M., Guzm{\'a}n F., Lykins M.~L., van Hoof P.~A.~M., Williams R.~J.~R., Abel N.~P., et al., 2017, RMxAA, 53, 385. doi:10.48550/arXiv.1705.10877

\bibitem[\protect\citeauthoryear{Fynbo et al.}{2013}]{Fynbo+13} Fynbo, J.~P.~U., Krogager, J.-K., Venemans, B., et al.\ 2013, ApJS, 204, 6

\bibitem[\protect\citeauthoryear{Gallagher et al.}{2007}]{Gallagher+07} Gallagher S.~C., Hines D.~C., Blaylock M., Priddey R.~S., Brandt W.~N., Egami E.~E., 2007, ApJ, 665, 157. doi:10.1086/519438

\bibitem[\protect\citeauthoryear{Ganguly et al.}{2007}]{Ganguly+07} Ganguly, R., Brotherton, M.~S., Cales, S., et al.\ 2007, ApJ, 665, 990

\bibitem[\protect\citeauthoryear{Gaskell}{2008}]{Gaskell08} Gaskell, C.~M. 2008, Revista Mexicana de Astronomia y Astrofisica Conference Series, 32, 1

\bibitem[\protect\citeauthoryear{Gaskell}{2017}]{Gaskell17} Gaskell C.~M., 2017, MNRAS, 467, 226. doi:10.1093/mnras/stx094

\bibitem[\protect\citeauthoryear{Gaskell et al.}{2023}]{Gaskell+23} Gaskell C.~M., Anderson F.~C., Birmingham S. {\'A}., Ghosh S., 2023, MNRAS, 519, 4082. doi:10.1093/mnras/stac3333

\bibitem[\protect\citeauthoryear{Gaskell \& Benker}{2007}]{Gaskell+Benker07} Gaskell, C.~M., \& Benker, A.~J.\ 2007, arXiv:0711.1013

\bibitem[\protect\citeauthoryear{Gaskell et al.}{2004}]{Gaskell+04} Gaskell, C.~M., Goosmann, R.~W., Antonucci, R.~R.~J., \& Whysong, D.~H.\ 2004, ApJ, 616, 147

\bibitem[\protect\citeauthoryear{Gaskell \& Harrington}{2018}]{Gaskell+Harrington18} Gaskell C.~M., Harrington P.~Z., 2018, MNRAS, 478, 1660. doi:10.1093/mnras/sty848

\bibitem[\protect\citeauthoryear{Gaskell, Klimek, \& Nazarova}{2007}]{Gaskell+07} Gaskell C.~M., Klimek E.~S., Nazarova L.~S., 2007, arXiv, arXiv:0711.1025. doi:10.48550/arXiv.0711.1025 (GKN)

\bibitem[\protect\citeauthoryear{Goosmann \& Gaskell}{2007}]{Goosmann+Gaskell07} Goosmann, R.~W., \& Gaskell, C.~M.\ 2007, A\&A, 465, 129

\bibitem[\protect\citeauthoryear{Hall et al.}{2002}]{Hall+02} Hall, P.~B., Anderson, S.~F., Strauss, M.~A., et al.\ 2002, ApJS, 141, 267

\bibitem[\protect\citeauthoryear{Hamann et al.}{2002}]{Hamann+02} Hamann, F., Sabra, B., Junkkarinen, V., Cohen, R., \& Shields, G.\ 2002, X-ray Spectroscopy of AGN with Chandra and XMM-Newton, p.~121 (= arXiv:astro-ph/0304564)

\bibitem[\protect\citeauthoryear{Heard \& Gaskell}{2016}]{Heard+Gaskell16} Heard, C.~Z.~P., \& Gaskell, C.~M.\ 2016, MNRAS, 461, 4227

\bibitem[\protect\citeauthoryear{Heard \& Gaskell}{2023}]{Heard+Gaskell23} Heard C.~Z.~P., Gaskell C.~M., 2023, MNRAS, 518, 418. doi:10.1093/mnras/stac2220

\bibitem[\protect\citeauthoryear{Herbst et al.}{2016}]{Herbst+16} Herbst, H., Hamann, F., Villforth, C., et al.\ 2016, American Astronomical Society Meeting Abstracts, 227, 104.03

\bibitem[\protect\citeauthoryear{Hopkins et al.}{2005}]{Hopkins+05} Hopkins P. F., Hernquist L., Martini P., Cox T. J., Robertson B., Di Matteo
T., Springel V., 2005, ApJ, 625, L71

\bibitem[\protect\citeauthoryear{Jaffarian \& Gaskell}{2020}]{Jaffarian+Gaskell20} Jaffarian G.~W., Gaskell C.~M., 2020, MNRAS, 493, 930. doi:10.1093/mnras/staa262

\bibitem[\protect\citeauthoryear{Jiang et al.}{2013}]{Jiang+13} Jiang, P., Zhou, H., Ji, T., et al.\ 2013, AJ, 145, 157

\bibitem[\protect\citeauthoryear{Kishimoto et al.}{2001}]{Kishimoto+01} Kishimoto, M., Antonucci, R., Cimatti, A., et al.\ 2001, ApJ, 547, 667

\bibitem[\protect\citeauthoryear{Kwan et al.}{2023}]{Kwan+23} Kwan E., Gaskell C.~M., Subramonian D.~G., Kamath A.~R., 2023, BAAS, 55, 177.84

\bibitem[\protect\citeauthoryear{Liu et al.}{2018}]{Liu+18} Liu H., Luo B., Brandt W.~N., Gallagher S.~C., Garmire G.~P., 2018, ApJ, 859, 113. doi:10.3847/1538-4357/aabe8d

\bibitem[\protect\citeauthoryear{Lynds}{1967}]{Lynds67} Lynds, C.~R.\ 1967, ApJ, 147, 396

\bibitem[\protect\citeauthoryear{Maddox \& Hewett}{2008}]{Maddox+Hewett08} Maddox, N., \& Hewett, P.~C.\ 2008, Memorie della Societ{\^a} Astronomica Italiana, 79, 1117

\bibitem[\protect\citeauthoryear{Maiolino et al.}{2001}]{Maiolino+01} Maiolino R., Marconi A., Salvati M., Risaliti G., Severgnini P., Oliva E., La Franca F., et al., 2001, A\&A, 365, 28. doi:10.1051/0004-6361:20000177

\bibitem[\protect\citeauthoryear{Mathews \& Ferland}{1987}]{Mathews+Ferland87} Mathews W.~G., Ferland G.~J., 1987, ApJ, 323, 456. doi:10.1086/165843

\bibitem[\protect\citeauthoryear{Meusinger et al.}{2005}]{Meusinger+05} Meusinger, H., Froebrich, D., Haas, M., et al.\ 2005, A\&Ap, 433, L25

\bibitem[\protect\citeauthoryear{Moravec et al.}{2016}]{Moravec+16} Moravec, E., Hamann, F., Capellupo, D.~M., et al.\ 2016, American Astronomical Society Meeting Abstracts, 227, 417.04

\bibitem[\protect\citeauthoryear{Nandy et al.}{1982}]{Nandy+82} Nandy, K., McLachlan, A., Thompson, G.~I., et al.\ 1982, MNRAS, 201, 1P

\bibitem[\protect\citeauthoryear{P{\^a}ris et al.}{2014}]{Paris+14} P{\^a}ris, I., Petitjean, P., Aubourg, {\'E}., et al.\ 2014, A\&Ap, 563, A54

\bibitem[\protect\citeauthoryear{Prevot et al.}{1984}]{Prevot+84} Prevot, M.~L., Lequeux, J., Prevot, L., Maurice, E., \&
Rocca-Volmerange, B.\ 1984, A\&Ap, 132, 389

\bibitem[\protect\citeauthoryear{Pu}{2014}]{Pu14} Pu, X.\ 2014, Ap\&SS, 349, 947

\bibitem[\protect\citeauthoryear{Reichard et al.}{2003}]{Reichard+03} Reichard, T.~A., Richards, G.~T., Hall, P.~B., et al.\ 2003, AJ, 126, 2594

\bibitem[\protect\citeauthoryear{Richards et al.}{2003}]{Richards+03}  Richards, G.~T., Hall, P.~B., Vanden Berk, D.~E., et al.\ 2003, AJ, 126, 1131

\bibitem[\protect\citeauthoryear{Rocca-Volmerange et al.}{1981}]{Rocca-Volmerange+81} Rocca-Volmerange, B., Prevot, L., Prevot-Burnichon, M.~L.,
Ferlet, R., \& Lequeux, J.\ 1981, A\&Ap, 99, L5

\bibitem[\protect\citeauthoryear{Scoville \& Norman}{1995}]{Scoville+Norman95} Scoville, N., \& Norman, C.\ 1995, ApJ, 451, 510

\bibitem[\protect\citeauthoryear{Sprayberry \& Foltz}{1992}]{Sprayberry+Foltz92} Sprayberry, D., \& Foltz, C.~B.\ 1992, ApJ, 390, 39

\bibitem[\protect\citeauthoryear{Trump et al.}{2006}]{Trump+06} Trump, J.~R., Hall, P.~B., Reichard, T.~A., et al.\ 2006, ApJS, 165, 1

\bibitem[\protect\citeauthoryear{Urrutia et al.}{2009}]{Urrutia+09} Urrutia, T., Becker, R.~H., White, R.~L., et al.\ 2009, ApJ, 698, 1095

\bibitem[\protect\citeauthoryear{Violino et al.}{2016}]{Violino+16} Violino, G., Coppin, K.~E.~K., Stevens, J.~A., et al.\ 2016, MNRAS, 457, 1371

\bibitem[\protect\citeauthoryear{Vito et al.}{2018}]{Vito+18} Vito F., Brandt W.~N., Luo B., Shemmer O., Vignali C., Gilli R., 2018, MNRAS, 479, 5335. doi:10.1093/mnras/sty1765

\bibitem[\protect\citeauthoryear{Weingartner \& Draine}{2001}]{Weingartner+Draine01} Weingartner, J.~C., \& Draine, B.~T.\ 2001, ApJ, 548, 296

\bibitem[\protect\citeauthoryear{Weymann et al.}{1991}]{Weymann+91} Weymann, R.~J., Morris, S.~L., Foltz, C.~B., \& Hewett, P.~C.\ 1991, ApJ, 373, 23

\bibitem[\protect\citeauthoryear{Yang et al.}{2022}]{Yang+22} Yang J., Fan X., Wang F., Lanzuisi G., Nanni R., Cappi M., Chartas G., et al., 2022, ApJL, 924, L25. doi:10.3847/2041-8213/ac45f2

\bibitem[\protect\citeauthoryear{Zhang et al.}{2014}]{Zhang+14} Zhang, S., Wang, H., Wang, T., et al.\ 2014, ApJ, 786, 42

\bibitem[\protect\citeauthoryear{Yamamoto \& Vansevi{\v c}ius}{1999}]{Yamoto+Vansevicius99} Yamamoto, T.~M., \& Vansevi{\v c}ius, V.\ 1999, PASJ, 51, 405

\bibitem[\protect\citeauthoryear{Yuan \& Wills}{2004}]{Yuan+Wills04} Yuan M.~J., Wills B.~J., 2004, AdSpR, 34, 2599. doi:10.1016/j.asr.2003.02.088

\bibitem[\protect\citeauthoryear{Zhou}{2006}]{Zhou+06} Zhou, H., Wang, T., Yuan, W., et al.\ 2006, ApJS, 166, 128

\bibitem[\protect\citeauthoryear{Zubko \& Laor}{2000}]{Zubko+Laor00} Zubko, V.~G., \& Laor, A.\ 2000, ApJS, 128, 245


\end{thebibliography}
\end{document}